\documentstyle[12pt,graphicx]{article}
\begin{document}
\title{Detecting very-high-frequency relic gravitational waves
             by  electromagnetic wave polarizations in a waveguide
             \thanks{supported by the CNSF, SRFDP, and CAS.}}

   \setcounter{page}{1}
\author{\small  Ming-Lei Tong and  Yang Zhang\thanks{yzh@ustc.edu.cn} \\
        \small Centre for Astrophysics,
        University of Science and Technology of
        China,Hefei 230026\\
      }

 \date{}

\def\be{\begin{equation}}
\def\ee{\end{equation}}
\def\ba{\begin{eqnarray}}
\def\ea{\end{eqnarray}}
\def\nn{\nonumber}
\def\bp{\mathbf{\Pi}}
\def\cf{\mathcal{F}}
\def\ch{\mathcal{H}}
\def\d{\mathrm{d}}
\def\x{\mathbf{x}}

\maketitle \baselineskip=19truept

\normalfont

\abstract{The polarization vector (PV) of an electromagnetic wave
(EW) will experience a rotation in a region of spacetime perturbed
by gravitational waves (GWs). Based on this idea, Cruise's group has
built an annular waveguide to detect GWs. We  give detailed
calculations of the rotations of the  polarization vector of
 an EW caused by incident GWs from
various directions and in various polarization states, and then
analyze the accumulative effects on the polarization vector when the
EW passes $n$ cycles along the annular waveguide. We reexamine the
feasibility and limitation of this method to detect GWs of high
frequency around  $100$ MHz, in particular, the relic gravitational
waves (RGWs). By comparing the spectrum of RGWs in the accelerating
universe with the detector sensitivity of the current waveguide, it
is found that the amplitude of the RGWs is too low to be detected by
the waveguide detectors currently running. Possible ways of
improvements on detection are discussed also.
 }

\section{ Introduction}

GW is one of important predictions of general relativity. Although
there has been an indirect evidence of GW radiation from the binary
pulsar B1913+16 \cite{taylor}, so far direct detection of GWs have
not been accomplished yet. GWs can have different frequencies
generated by various kinds of sources. Currently, besides the
conventional method of cryogenic resonant bar \cite{bar}, a number
of detectors using new techniques have been running or under
construction aiming at direct signals of GWs. For a frequency range
$1\sim 10^4$ Hz, the method of ground-based laser interferometers
applies, such as LIGO \cite{Abramovici}, Virgo \cite{Bradaschia},
and TAMA \cite{tama}, . For a lower frequency range $10^{-4}\sim 1$
Hz, the space-based laser interferometers can be used, such as LISA
\cite{Jafry} under planning. For much lower frequencies $\sim
10^{-18}$ Hz, detections of CMB polarization of ``magnetic''  type
would also give direct evidence of GWs \cite{Bpol}. There have also
been attempts to detect GWs of very high frequencies from MHz to
GHz, employing various techniques, such as laser beam \cite{li}. One
interesting method proposed by Cruise uses linearly polarized EWs
\cite{cruise2, cruise3}. When GWs pass through the region of
waveguide, the direction of PV  of EWs will generally experience a
rotation \cite{cruise2}. A prototype gravitational waves detector
has been built by Cruise's group \cite{cruise3}, which consists
mainly of one, or several, annular waveguide of a shape of torus. As
a merit of this method, depending upon the size of the waveguide,
GWs in a very high frequency range $ 10^6\sim 10^9$ Hz can be
detected, which is not covered by the laser interferometer method.
Note that the GWs in the frequency range $ 10^6\sim 10^9$ Hz are
generally not generated by usual astrophysical processes, such as
binary neutron stars, binary black holes, merging of neutron stars
or black holes, and collapse of stars \cite{grishchuk5}
\cite{zhang2}. However, the background of RGWs has a spectrum
stretching over a whole range of $10^{-18}\sim 10 ^{11}$ Hz
\cite{Grishchuk,Zhang}. Depending on the frequency ranges, its
different portions can be detected by different method. For
instance, the very low frequency range $10^{-18}\sim 10 ^{-16}$ Hz
can be detected by the curl type of polarization in CMB \cite{Bpol},
the low frequency range $10^{-3}\sim 10 ^{-2}$ Hz can be detected by
LISA, the mediate frequency range $10^{2}\sim 10 ^{3}$ Hz is covered
by LIGO, and the very high frequency range $10^{6}\sim 10 ^{9}$ Hz
can be the detection object of Cruise's EWs polarization method.
Therefore, one of the main object of detection by the annular
waveguide is the very high frequency RGWs. The detection of high
frequency RGWs from  MHz to GHz is in complimentary to the usual
detectors working in the range of $10^{-4}\sim10^4$ Hz. RGWs  is a
stochastic background that are generated by the inflationary
expansion of the early Universe \cite{Grishchuk,Zhang,giovamini,
zhaozhang, Miao}, and its spectrum depends sensitively on the
inflationary  and the subsequent reheating stages. Besides, the
currently accelerating expansion also affects both the shape and the
amplitude of the RGW spectrum \cite{Grishchuk,Zhang,Miao}. RGWs
carry take a valuable information about the  Universe, therefore,
their detection is much desired and will provide a new  window of
astronomy.

In this paper we give a comprehensive study  of
the rotations of PV of EWs in a conducting torus caused by incident GWs,
and explore the feasibility and limitation
of Cruise's method of detecting GWs by
polarized EWs in the annular waveguide.
Firstly, we briefly review the RGWs in the
currently accelerating universe.
Secondly, we shall present detailed calculations of rotations of
the PV of EWs in the waveguide
caused by the incoming GWs from from various directions
and in various polarization states,
thereby we analyze the multiple-cycling accumulating effect
and the resonance when the circling frequency of EWs is
nearly equal to that of GWs.
Thirdly, we shall examine the possible detection of
the RGWs by the annular waveguide system around 100 MHz,
comparing the predicted spectrum of RGWs
in the accelerating Universe
with the sensitivity of the detector  \cite{cruise3}.
Finally, we give the conclusions and possible ways of
improvements for detection.

\section { Relic gravitational waves   }

\begin{figure}
\centerline{\includegraphics[width=8cm]{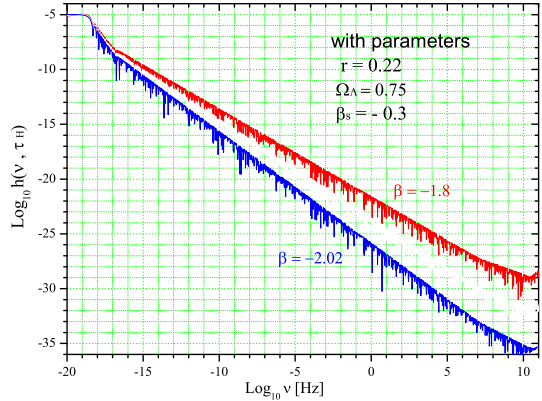}}
\caption{\label{fig1}
The spectrum $h(\nu, \eta_H)$ of RGW
in the accelerating universe. }
\end{figure}

In an expanding universe RGWs can be regarded as
small perturbations to the Robertson-Walker metric,
 \be \label{RW}
 {\d}s^2=a^2(\tau)[\,-{\d}\tau^2  + (\delta_{ij}+h_{ij})\,{\d}x^i {\d}x^j],
 \ee
where $a(\tau)$ is the scale  factor,
$\tau$ is the conformal time, and $h_{ij}$ is transverse-traceless
\be
 \partial_i\,h^{ij}=0, \qquad \  \delta^{ij}\,h_{ij}=0,
\ee
representing RGWs.
Among six components $h_{ij}$ there are only two independent
(two polarization states).
Generally, $|h_{ij}|\ll 1$.
The wave equation for RGWs is
 \be \label{wave equation}
 \partial_\mu\,(\sqrt{-g}\,\partial^\mu\,h_{ij}({\x},\tau))=0.
 \ee
The solution $h_{ij}$ of Eq.({\ref{wave equation}})
and the spectrum $h(\nu,\tau_{H})$ defined via
\be
<h^{ij}h_{ij}> = \int_0^\infty h^2(k,\tau_H)\frac{{\d}k}{k}
\ee
have been given for an accelerating universe \cite{Zhang,Miao,Grishchuk2}.
 Fig.\ref{fig1} plots $h(\nu,\tau_{H})$,
which depends on the accelerating parameter $\gamma$, the inflation
parameter $\beta$, the reheating parameter $\beta_s$,
the tensor/scalar ratio $r$,
and the redshift $z_E$ at the time $\tau_E$
of equality of dark energy and matter is given by
\be\label{redshift}
1+z_E=\frac{a(\tau_H)}{a(\tau_E)}
 \simeq  (\frac{\Omega_\Lambda}{\Omega_m})^{\frac{1}{3}}.
\ee
Since the annular waveguide is to detect RGWs of
frequencies $\sim 10^8 $ Hz,
we quote the analytic approximate spectrum in this range \cite{Zhang}:
 \be\label{spectrum}
h(k,\tau_H)\approx
A_0(\frac{k_s}{k_H})^{\beta_s}\frac{k_H}{k_2}(\frac{k}
{k_H})^{\beta-\beta_s+1}\frac{1}{(1+z_E)^{3+\epsilon}},
 \ee
where  $k$ is the comoving wavenumber related to
the physical frequency by $\nu=\frac{k}{2\pi a(\tau_H)}$,
$A_0$ is a constant determined by the CMB anisotropies
\cite{Bpol,Zhang,Grishchuk2,Spergel},
$\epsilon\equiv(1+\beta)(1-\gamma)/\gamma$ is a small parameter,
and $k_H=2\pi \gamma$ and $k_s\simeq10^{26}k_H$.
Note that the RGWs $h_{ij}$ described above exist everywhere and
all the time in the Universe.
We may simply say that the Universe is filled
with a stochastic background,
consisting of all the modes of different
wave-vector $k^{\mu}$=$(k^0,k^1,k^2,k^3)$.
So the RGWs serve as an object for GWs detections.

In the frequency range $\sim 100 $ MHz
for the waveguide detector,
RGWs can be approximated as plane waves.
A beam of  monochromatic plane GWs with a wave-vector
 can be generally written as the following form \cite{mtw}
 \be \label{metric}
h_{ij}=Re\{A_{ij}e^{i\phi}\},
 \ee
where $A_{ij}$ represents the amplitude and
$\phi$  is the  phase of GWs,
\be  \label{phi}
 \phi=k_\mu x^\mu=
g_{\mu\nu}k^\mu x^\nu,
 \ee
with  $x^\mu$ being the point of spacetime that the waves pass.

\section{The Annular Waveguide }

\begin{figure}
\centerline{\scalebox{1.2}[1.2]{\includegraphics[width=8cm]{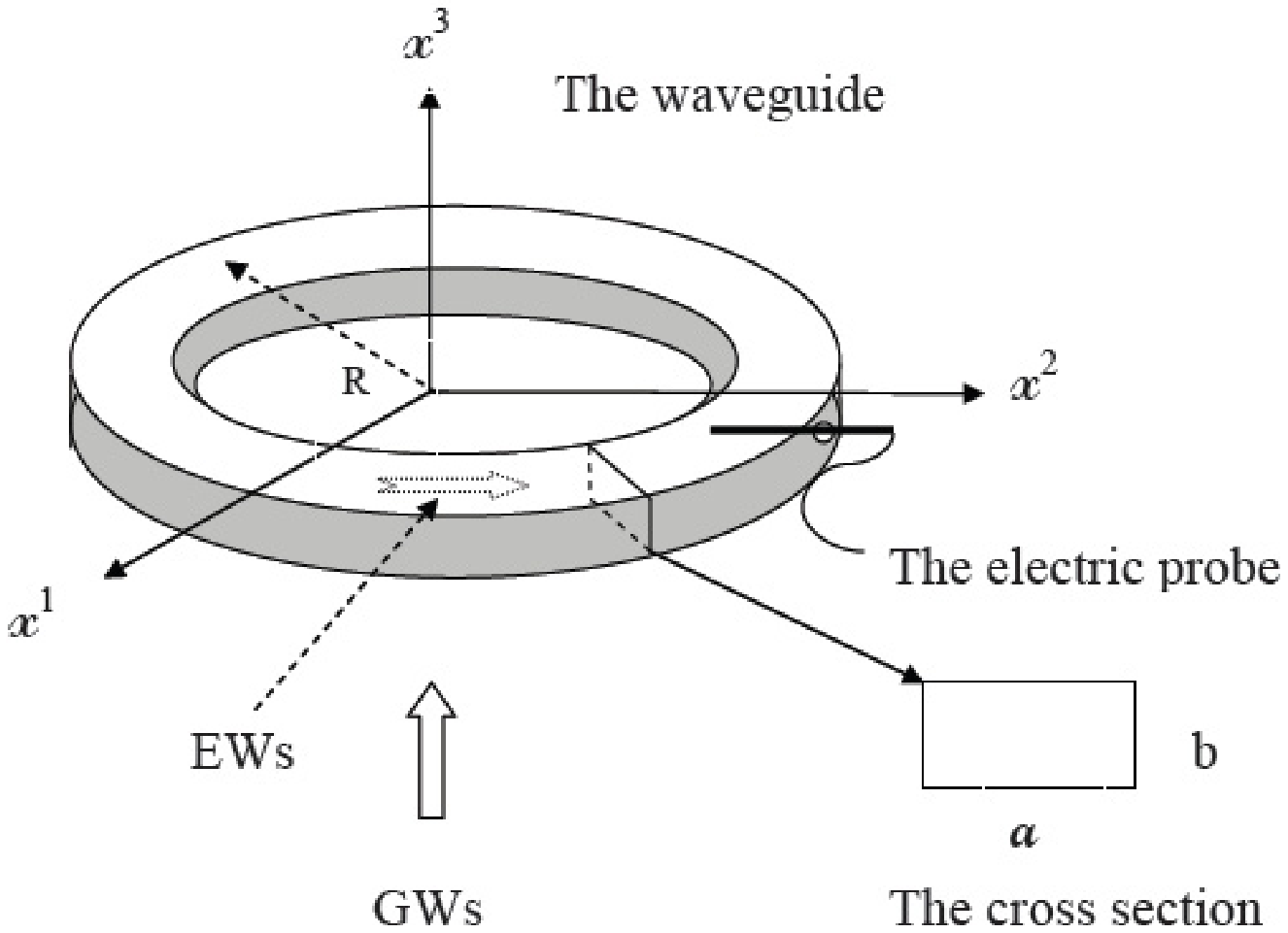}}}
 \caption{\label{fig2}
A sketch map of the annular waveguide.
The cross section of the waveguide is
rectangle with  $a$ and  $b$ being much less than the radius $R$.
EWs travel inside
 the waveguide and GWs propagate along the $x^3$-axis.}
\end{figure}

Consider an annular waveguide of shape of a torus,
as shown Fig.\ref{fig2}.
Its  radius  is $R$, and the cross section is a rectangle with
sides $a>b$,  both being  much less than $R$,
say, $a,b \sim  1$ cm, and $R\sim 1$ m.
(Note that the waveguide actually employed
by Cruise's group \cite{cruise3}
actually  has a shape of rectangle, instead of a torus.
For simplicity of analysis,
here we consider a torus since the working mechanism is the same.)
Inside the torus,  one can input a beam of linearly polarized
EW propagating around the toroidal loop,
and the beam consists of a ${\mathrm{TE}_{10}}$
mode (transverse electric field)
with the electric field $\bf E$ pointing along the $x^3-$axis.
The EWs are a microwave, say,  with a wavelength $\lambda_e \sim  1$ mm
and a frequency $\nu_e =c/\lambda_e\sim 10^{11}$ Hz.
The guided EWs of ${\mathrm{TE}_{10}}$ mode
in the torus travel around the loop at a group speed  \cite{Li},
\be
v =c\sqrt{1-(\frac{\lambda_e}{ 2a})^2},
 \ee
where $c$ is the speed of light,
and $v$ is very close to $c$.
For instance, for  $\lambda_e\sim 1$mm and  $a= 1$cm,
the difference between the two velocities
is $\sim (\lambda_e/2a)^2/2\sim 10^{-3}$.
As is well-known, for a TE$_{10}$ mode to exist in the waveguide,
one has to $\lambda_e\le 2a$.
The angular velocity of EWs cycling around the loop is then
\be
\omega_0 = \frac{v}{R} \simeq \frac{c}{R}.
\ee
As will be seen later,
when the angular frequency $\omega$ of
the incident GWs is very close to the $\omega_0$,
i.e., at the resonant condition,
the detector responds most sensitively to the GWs.
Therefore, such a device of given radius $R$
will primarily detect  GWs of a resonant frequency around
 \be \label{nu-g}
 \nu_g \simeq \frac{c}{2\pi R}.
 \ee
For example,  if the radius is $R=1$ m, then the frequency of
GWs to be detected is $\nu_g\simeq 5\times10^7$ Hz,
some two orders smaller than $\nu_e$.
As a merit, by adjusting the size $R$,
the frequency of GWs to be detected can  vary accordingly.
For the 4-dim spacetime, one can choose
a coordinate system $\{x^{\mu}\}$
with $\mu=0,1,2,3$ and $x^0\equiv ct$,
such that  the waveguide lies on the $(x^1,x^2)$ plane
as shown in Fig.\ref{fig2}.
Note that the geometric size $R$ of waveguide
is negligibly small in comparison
with the Hubble's radius $\sim c/H$,
so  that the effect of cosmic expansion
on the torus can be totally neglected.

Let us consider a beam of GWs passing through  the detector.
Assume that the wavelength $\lambda_g$ of the GWs is much longer than
the wavelength $\lambda_e$ of the EWs in the waveguide,
 i.e.,
$\lambda_g \gg \lambda_e$,
so that the geometric optics approximation applies
in describing the EWs \cite{cruise2,mtw}.
In fact, this assumption on the incident GWs is automatically satisfied
if the GWs satisfy the resonant condition.
The PV of the linearly polarized EWs can be described by a 4-vector,
$\Pi^{\mu}=( \Pi^0, \Pi^1, \Pi^2, \Pi^3)$,
which is real and normal to the wave vector
$P^\mu$ of the EWs
\be  \label{vertical}
 \Pi_{\mu}  P^{\mu}  =0,
\ee
and satisfies the normalized condition \cite{mtw, cruise2}
\be  \label{unity}
   \Pi_{\mu} \Pi^{\mu}=1.
\ee Eq.(\ref{vertical}) tells that one can add a multiple of
$P^{\mu}$ to $\Pi^{\mu}$ without affecting any physical measurements
\cite{mtw}, since $P^\mu$ is a null vector with $P_\mu P^\mu=0$.
Suppose that the EWs are propagating along the $x^1-$axis with the
wave vector $P^\mu=(P^0, P^1,0,0)$, which satisfies $P_\mu P^\mu=0$.
Then, by Eq.(\ref{vertical}), the PV of EWs can be generally written
as $\Pi^\mu = (\kappa P^0, \kappa P^1,\Pi^2, \Pi^3)$, where $\kappa$
is an arbitrary constant. Then,   Eq.(\ref{unity}) leads to \be
|\Pi^2|^2+|\Pi^3|^2=1. \ee Since initially the electric field
$\mathbf{E}$ of the EWs inside the torus is set to be along the
$x^3-$  axis
 and $\Pi^i$ is, by definition,  in the direction of $\mathbf{E}$,
so the initial PV is
\be \label{initialp}
\Pi^{\mu}=(\,0, 0, 0, 1),
\ee
i.e., initially the PV has a vanishing  $\Pi^2=0$ component.

However, the presence of  GWs will cause a rotation of
the $\Pi^\mu$ about the direction of propagation,
generating a non-vanishing $\Pi^2\ne 0$
i.e., a component $E^2\ne 0$ of the electric field $\mathbf{E}$
of the EWs in the waveguide.
One puts an electric field probe inside the waveguide
at the intersection of the $x^2-$ axis and the torus.
The probe is on the line of $x^2$ axis,
so that it can probe the non-vanishing electric field $E^2$
due to the rotation of $\bf E$ caused by the GWs \cite{cruise3}.
The GWs induce  an electric voltage on the electric probe
$V = E_0 \alpha \, l\, \sin{(2\pi \nu_e  t)}$,
where $E_0$ is the  ${\mathrm{TE}}_{10}$ mode electric field in the
waveguide, $l$ is the length of the conducting probe.

In the geometric optics approximation,
the motion of $\Pi^{\mu}$ is described as being
 parallel-propagating along the rays of Ews with the equation
 \be \label{parallel}
 \frac{d\Pi^\mu}{ds}+\Gamma_{\nu\sigma}^\mu\Pi^\nu\frac{dx^\sigma}{ds}=0,
 \ee
where $s$ is an affine parameter, which  can be chosen to be $s=t/T_0$
with $T_0= 2\pi /\omega_0$
being the period of the EWs travelling around the torus.
Note that when $s$ goes from $0$ to $1$,
the EWs go one cycle around the torus.
Since  $a, b\ll R$,
one can view the EWs in the waveguide as travelling
along the $1-$dimensional loop path
 \be\label{x}
x^{\mu}= R\left(\frac{2\pi c s  }{v}, -\sin{2\pi s}, \cos{2\pi s},
 0\right),
\ee
where $v$ is the group speed of the EWs.

With the initial setup of the polarization of EWs in the torus,
we only need to consider the component
$\Pi^2$  of the polarization in the following.

\section{Change of $\Pi^2$}

Even the setup of the waveguide detector is fixed in laboratory,
GWs propagating in space may come in any direction randomly.
Therefore, we need to determine the rotation of polarization of EWs
caused by GWs travelling along the directions  $x^i$, $i=1,2,3$,
respectively.

\begin{center}
{\bf  A.  GWs travelling  along the positive $x^3$-axis }
\end{center}

Consider a beam  of monochromatic plane GWs travelling
along the positive $x^3$-axis
with a wave vector $k^\mu=(2\pi/\lambda_g,0,0,2\pi/\lambda_g)$.
As the GWs just pass the annular waveguide whose position is
given by Eq.(\ref{x}),
then substituting it into Eq.(\ref{phi})
yields the phase of the GWs at the point inside the annular waveguide
\be \label{phi3}
   \phi=-2\pi s\, \omega/\omega_0.
\ee
Here $\omega =2\pi c/\lambda_g$ represents the angular frequency of GWs,
and $\omega_0 = (2\pi/T_0)$ is
the cycling angular frequency of EWs around the torus.
As mentioned before,
one can take the flat spacetime slightly perturbed by GWs
to represent the local region of the waveguide.
In the transverse traceless (TT) gauge,
the metric tensor can be written as
\[
 g_{\mu\nu}=\eta_{\mu\nu}+h_{\mu\nu}=
 \left(\begin{array}{cccc}
-1 & 0 & 0 & 0\\0 & 1+h_{\oplus} & h_{\otimes}& 0\\0 & h_{\otimes} &
1-h_{\oplus} & 0\\0 & 0 & 0 & 1\end{array}\right),
\]
 and
 \[
g^{\mu\nu}=\eta^{\mu\nu}-h^{\mu\nu}=\left(\begin{array}{cccc} -1 & 0
& 0 & 0\\0 & 1-h_{\oplus} & -h_{\otimes}& 0\\0 & -h_{\otimes} &
1+h_{\oplus} & 0\\0 & 0 & 0 & 1\end{array}\right),
 \]
where  $h_\oplus$ and $h_{\otimes}$ denote the $+$ and
 $\times$ modes of polarization of GWs,  respectively.
In general, these two modes of GWs may be not coherent,
i.e. their phases are random and independent,
a situation similar to the natural light of EWs.
If the $+$ and $\times$ modes have the same phase $\phi$,
which  is called as the linearly polarized  GWs \cite{mtw},
then by Eq.(\ref{metric})
one has
\be \label{metric1}
     h_\oplus=A_\oplus\cos{\phi}, \,\,\,\,
     h_\otimes=A_\otimes\cos{\phi},
\ee
where $A_\oplus$ and $A_\otimes$ are real numbers.
This case of a linearly polarized  GWs will be
discussed in the following,  otherwise we give a clear indication.

As can be checked,
the change in $\Pi^3$ due to GWs  is of  order of $|h_{ij}|^2$,
so in the subsequent calculation, $\Pi^3=1$ is assumed.
To calculate the change of $\Pi^2$ up to the linear order of $h_{ij}$,
one needs  the following  Christoffel components
\ba \label{gamma}
&&\Gamma_{31}^2=-\pi A_{\otimes}\sin{\phi}/\lambda_g \nonumber,\\
&&\Gamma_{32}^2=\pi A_{\oplus}\sin{\phi}/\lambda_g,
\ea
other components are either zero or of order $|h_{ij}|^2$,
having no contributions.
Integrating Eq.(\ref{parallel})
gives the expression of the change in $\Pi^2$
around one circle of the torus
\be  \label{increment}
\Delta\Pi^2=
 \int_{0}^{1}\frac{d\Pi^2}{ds}d s=- \int_{0}^{1}
 \left(\Gamma_{31}^{2} \Pi^{3} \frac{dx^1}{ds}
 +\Gamma_{32}^{2} \Pi^{3} \frac{dx^2}{ds}\right) ds.
 \ee
Substituting Eqs.(\ref{x}) and (\ref{gamma}) into the integration,
one has
\be \label{increment1}
 \Delta\Pi^2=\frac{2\pi^2 R}{\lambda_g}
 \int_{0}^{1}\left(A_{\otimes}\sin{\bigg(2\pi
 s\frac{\omega}{\omega_0}\bigg)}\cos{2\pi s}-A_{\oplus}\sin{\bigg(2\pi
 s\frac{\omega}{\omega_0}\bigg)}\sin{2\pi s}\right) ds.
\ee
Carrying out integration yields the result
\be\label{Delta2}
 \Delta\Pi^2=\frac{A_{\otimes}}{2}\left(1-\cos{(2\pi
\varpi)}\right)\frac{\varpi^2}{\varpi^2-1}
-\frac{A_{\oplus}}{2}\sin{(2\pi
\varpi)}
\frac{\varpi}{\varpi^2-1},
 \ee
where  $\varpi\equiv \omega/\omega_0$.
So the change of $\Pi^2$ depends on $\omega$.

Let us see what a value $\Delta\Pi^2$
will  take when the cycling angular frequency of EWs is equal to
the angular frequency of GWs,
\be
\omega_0  =\omega,
\ee
called the resonant condition.
Taking the limit $\varpi\rightarrow 1$ in Eq.(\ref{Delta2}) yields
a constant value
\be  \label{Delta1}
\Delta\Pi^2= -\frac{\pi A_{\oplus}}{2},
\ee
which has only contribution from the $+$ mode.
This is the known result by Cruise \cite{cruise2}.

Let us discuss  other special cases of Eq.(\ref{Delta2}).

\begin{figure}
\centerline{\includegraphics[width=8cm]{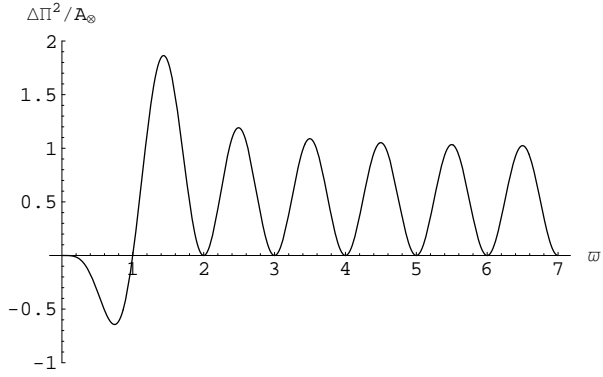}}
\caption{\label{fig3} The  $\Delta\Pi^2$ as an oscillating  function of
$\varpi$ when $A_{\oplus}=0$.
$\Delta\Pi^2$ has a
maximal value of $1.864A_{\otimes}$ at $\varpi=1.434$ and a
minimal value of $-0.643A_{\otimes}$ at $\varpi=0.743$.
Also notice that:
(1) $\Delta\Pi^2=A_{\otimes}$, for $\varpi\gg 1$
and $\varpi$ equals half integer;
(2)  $\Delta\Pi^2=0$, for  $\varpi$ equals integer;
(3) $\Delta\Pi^2=0$, for $\varpi\rightarrow 0$.}
\end{figure}

  (1) If the GWs are given such that $A_{\oplus}=0$, i.e.,
there is only  the $\times$ mode,
\be
   \Delta\Pi^2=\frac{A_{\otimes}}{2}
   \left(1-\cos{(2\pi\varpi)}\right)
   \frac{\varpi^2}{\varpi^2-1},
 \ee
which is plotted in  Fig. \ref{fig3} as a function of $\varpi$.
It shows that $ \Delta\Pi^2$ can be both
positive and negative, depending on $\varpi$. A maximum value of
$\Delta\Pi^2$ is achieved at $\varpi \simeq  1.434$.

\begin{figure}
\centerline{\includegraphics[width=8cm]{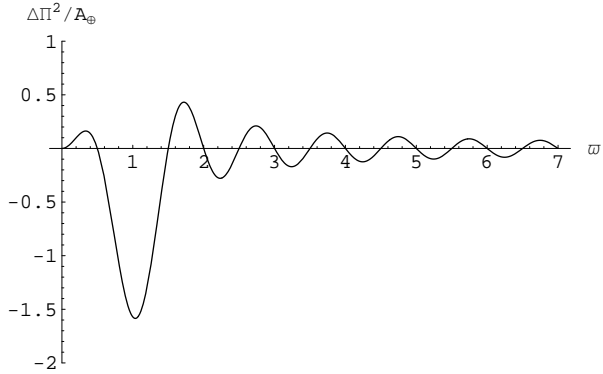}}
\caption{\label{fig4} The  $\Delta\Pi^2$ as a function of
$\varpi$ when $A_{\otimes}=0$.
$\Delta\Pi^2$ has a  minimum
$-1.585A_\oplus$ at $\varpi=1.036$.}
\end{figure}

(2) If  the GWs are given such that $A_{\otimes}=0$,
i.e., there is only the $+$ mode,
\be
   \Delta\Pi^2=-\frac{A_{\oplus}}{2}\sin{(2\pi
\varpi)} \frac{\varpi}{\varpi^2-1},
 \ee
which is shown in Fig. \ref{fig4}.
A minimum value of $\Delta\Pi^2$ is achieved at $\varpi \simeq 1.036$.

\begin{figure}
\centerline{\includegraphics[width=8cm]{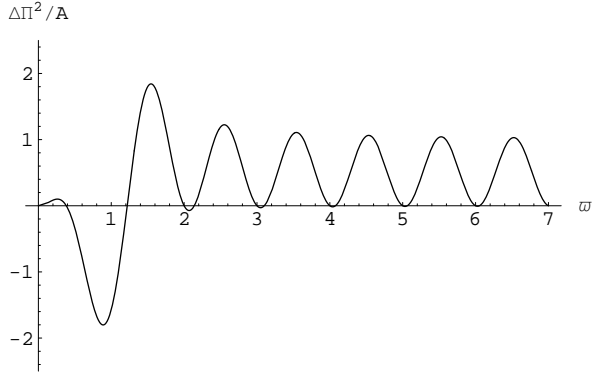}}
\caption{\label{fig5} The  $\Delta\Pi^2$ as a function of
$\varpi$ when $A_{\otimes}=A_{\oplus}= A$.
$\Delta\Pi^2$ has a maximum $1.842$ at $\varpi=1.546$ and a
minimum $-1.802$ at $\varpi=0.889$.}
\end{figure}

(3) If $A_{\otimes}= A_{\oplus}= A$, where  $A$ is real,
 as is likely the case for relic
gravitational waves, then one gets \be \Delta\Pi^2= \frac{A}{2}
(\varpi-\varpi
\cos(2\pi\varpi)-\sin(2\pi\varpi))\frac{\varpi}{\varpi^2-1}, \ee
which is shown in Fig.\ref{fig5} that is the compound of
Fig. \ref{fig3} and Fig. \ref{fig4}.

\begin{figure}
\centerline{\includegraphics[width=8cm]{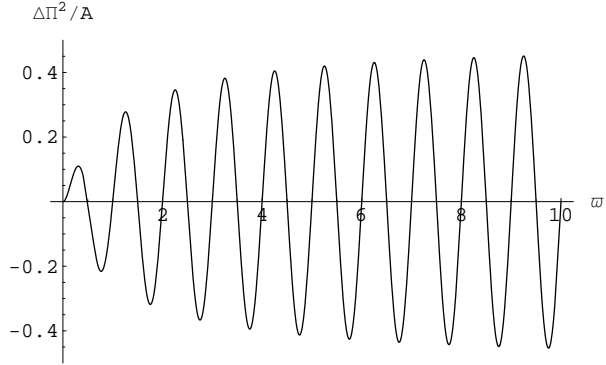}}
 \caption{\label{fig6} The relation between $\Delta\Pi^2$ and
$\varpi$ for circularly  polarized GWs.
The value of $\Delta\Pi^2$ will always less than  $A/2$. }
\end{figure}

Instead of a linearly polarized GWs in Eq.(\ref{metric1}),
 consider  the case of
circularly  polarized GWs with $A_\oplus=iA_\otimes=A$,
 \be \label{metric2}
     h_\oplus=A\cos{\phi}, \,\,\,\,\,
     h_\otimes=A\sin{\phi}.
 \ee
By similar calculations, one has the relevant Christoffel components
 \ba \label{gamma2}
&&\Gamma_{31}^2=\pi A\cos{\phi}/\lambda_g \nonumber,\\
&&\Gamma_{32}^2=\pi A\sin{\phi}/\lambda_g,
 \ea
Integrating  Eq.(\ref{increment}) yields
 \be \label{circulargw}
 \Delta\Pi^2 =\frac{A\varpi\sin{(2\pi \varpi)}}{2(1+\varpi)},
 \ee
which is shown in Fig.\ref{fig6}.

\begin{center}
{\bf B. GWs travelling along the positive $x^1$-axis }
\end{center}

Different from the above,
consider a plane GWs travelling along the positive $x^1$-axis.
The wave vector  is  $k^\mu=2\pi/\lambda_g(1,1,0,0)$.
Then,  by Eqs.(\ref{phi}) and (\ref{x}),
the phase of GW in the torus is
\be \label{phi2}
\phi=-2\pi x^0/\lambda_{g}+2\pi x^1/\lambda_{g}
          = -(2\pi s+\sin{(2\pi s)})\omega/\omega_0 .
\ee
The metric  is now
\[
 g_{\mu\nu}=\eta_{\mu\nu}+h_{\mu\nu}=
 \left(\begin{array}{cccc}
-1 & 0 & 0 & 0\\0 & 1 & 0 & 0\\0 & 0 & 1+h_{\oplus} & h_{\otimes}\\0
& 0 & h_{\otimes} & 1-h_{\oplus}\end{array}\right).
\]
Similar calculations give the relevant Christoffel components
\ba \label{gamma1}
&&\Gamma_{30}^2=\pi A_{\otimes}\sin{\phi}/\lambda_g \nonumber,\\
&&\Gamma_{31}^2=-\pi A_{\otimes}\sin{\phi}/\lambda_g,
 \ea
and the change in $\Pi^2$ around one circuit of the path
\be\label{increment3}
 \Delta\Pi^2=\frac{2\pi^2 r A_{\otimes}}{\lambda_g}
 \int_{0}^{1}
 (1+\cos{(2\pi s)})\sin{\bigg(\frac{\omega}{\omega_0}(2\pi
 s+\sin{(2\pi s)})\bigg)} ds.
 \ee
Integrating Eq.(\ref{increment3}) gives rise to
\be\label{increment4}
  \Delta\Pi^2=A_{\otimes}\sin^2{(\pi \varpi)}
\ee
which  is  oscillates between $A_{\otimes}$ and  $0$.
Note that the $A_\oplus$ has no contribution.

In the case of circularly polarized GWs,
 Eqs.(\ref{gamma1}) and (\ref{increment3}) should be replaced by
 \ba \label{gamma3}
&&\Gamma_{30}^2=-\pi A\cos{\phi}/\lambda_g \nonumber,\\
&&\Gamma_{31}^2=\pi A\cos{\phi}/\lambda_g,
 \ea
and
 \be\label{increment5}
 \Delta\Pi^2=\frac{2\pi^2 r A}{\lambda_g}
 \int_{0}^{1}
 (1+\cos{(2\pi s)})\cos{\bigg(\frac{\omega}{\omega_0}(2\pi
 s+\sin{(2\pi s)})\bigg)}ds,
 \ee
and one has
 \be\label{increment6}
\Delta\Pi^2=\frac{A}{2} \sin{(2\pi \varpi)},
 \ee
oscillating between $A/2$ and  $-A/2$.

\begin{center}
{\bf c.  GWs travelling  along the positive $x^2$-axis }
\end{center}

When the plan GWs  travel along the positive $x^2$-axis,
the metric tensor of spacetime is
\[
 g_{\mu\nu}=\eta_{\mu\nu}+h_{\mu\nu}=
 \left(\begin{array}{cccc}
-1 & 0 & 0 & 0\\0 & 1+h_{\oplus} & 0 & h_{\otimes}\\0 & 0 & 1 & 0 \\
0 & h_{\otimes} & 0 & 1-h_{\oplus}\end{array}\right).
\]
Similar calculations
show that the relevant Christoffel components are $0$, and thus
\be\label{5}
 \Delta\Pi^2=0.
 \ee
Thus, the GWs travelling along
 $x^2$-axis will not change $\Pi^2$.
Therefore, to avoid a null result of detection
in case of an incident GWs in the $x^2$ direction,
one may put two probes with $90^0$ separation along the
annular waveguide.

\section{ Accumulative  effect }

When the EWs pass $n$ cycles along the annular waveguide,
the change of $\Pi^2$ may be accumulative.
This is of practical significance in actual detections.
We need only consider GWs along the $x^3-$ and the $x^1-$
directions.

\begin{figure}
\centerline{\includegraphics[width=8cm]{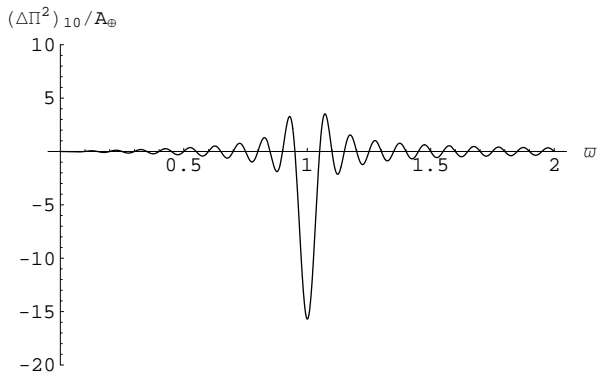}}
\caption{\label{fig7}
The case of $A_{\otimes}=0$ and $n=10$.
$(\Delta\Pi^2)_n$ has a  minimal  $-15.071A_{\oplus}$
at $\varpi=1.00038$. }
\end{figure}

Firstly, for the linearly polarized
 incident GWs in the $x^3-$ direction,
integrating Eq.(\ref{increment1}) form $0$ to $n$ gives
the change of $\Pi^2$ for  $n$ cycles
 \be\label{Deltan1}
  (\Delta\Pi^2)_{n}=\frac{A_{\otimes}}{2}
  \left(1-\cos{(2\pi n \varpi)}\right)\frac{\varpi^2}{\varpi^2-1}
   -\frac{A_{\oplus}}{2}\sin{(2\pi n \varpi)}\frac{\varpi}{\varpi^2-1}.
 \ee
In the special case $A_{\otimes}=0$,
Fig.\ref{fig7} gives a plot of $(\Delta \Pi^2)_n$ for  $n=10$.
In contrast with Fig.\ref{fig6} for $n=1$,
$(\Delta\Pi^2)_n$ is now sharply peaked
at $  \varpi \simeq 1$ with a much larger amplitude,
as a prominent feature.
As given in Table \ref{table2},
under the resonance condition $\varpi\rightarrow 1$,
the amplitude of $(\Delta\Pi^2)_n$ increases linearly with $n$.
In fact, this linearly-increasing
amplitude $\Delta\Pi^2_{\mathrm{min}}$ at very large $n$
is also obtained
by taking the resonance limit $\varpi\rightarrow 1$ of
Eq.(\ref{Deltan1}),
 yielding
 \be \label{Deltan2}
    (\Delta\Pi^2)_{n}=-\frac{n\pi A_\oplus}{2},
 \ee
which is in accord with the result obtained by
Cruise~\cite{cruise2}.

\begin{table}
 \caption{\label{table2}
 The case $A_\otimes=0$.
 The amplitude of $(\Delta\Pi^2)_n$ increases linearly with $n$,
  as $\varpi\rightarrow 1$.}
\begin{center}
 \begin{tabular}{|l|ll|}\hline
 n&$\varpi_{\rm{min}}$&$\Delta\Pi^2_{\rm{min}}/A_\oplus$ \\
 \hline
$1$&$1.036$&$-1.585$\\
\hline
$10$&$1.00038$&$-15.71$\\
\hline
 $100$&$\sim1$&$-157.08$\\
 \hline
 $1000$&$\sim1$&$-1570.8$\\
 \hline
 $2000$&$\sim1$&$-3141.6$\\
 \hline
 $10000$&$\sim1$&$-15708$\\
 \hline
\end{tabular}
\end{center}
\end{table}

\begin{figure}
\centerline{\includegraphics[width=8cm]{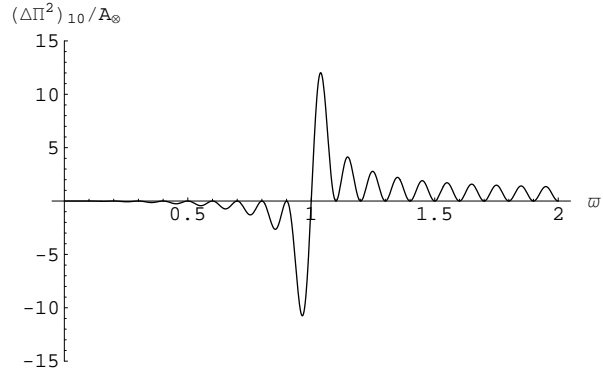}}
\caption{\label{fig8}
The case of $A_{\oplus}=0$ and $n=10$.
$(\Delta\Pi^2)_n$ has a maximum $\Delta\Pi^2_{max}=12.027A_{\otimes}$
   at $\varpi=1.038$,
and a minimal $\Delta\Pi^2_{min}=-10.761A_{\otimes}$ at $\varpi=0.964$. }
\end{figure}

\begin{figure}
\centerline{\includegraphics[width=8cm]{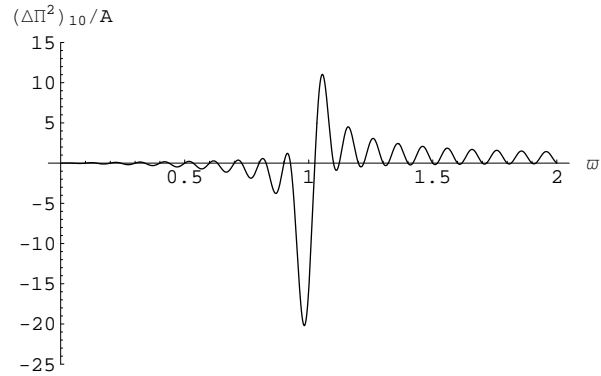}}
\caption{\label{fig9}
The case of $A_{\otimes}= A_{\oplus}=A$ and $n=10$.
$(\Delta\Pi^2)_n$ has a maximum $\Delta\Pi^2_{max}=11.05A$
         at $\varpi=1.055$,
and a minimal $\Delta\Pi^2_{min}=-20.214A$ at $\varpi=0.982$. }
\end{figure}

The special cases of $A_\oplus=0$  and of $A_\oplus=A_\otimes$
are quite similar to each other.
$(\Delta\Pi^2)_n$ has, for each given $n$,
both a sharp maximum $\Delta\Pi^2_{max}$ at $\varpi_{max}>1$
and a sharp minimum $\Delta\Pi^2_{min}$ at $\varpi_{min}<1$.
As $n\rightarrow \infty$,
the amplitudes $\Delta\Pi^2_{max}$ and $\Delta\Pi^2_{min}$
increase with $n$  approximately linearly,
and their locations $\varpi_{max}$ and $\varpi_{min}$
approach to $1$ from either side, respectively.
Fig.\ref{fig8} and Fig.\ref{fig9} give the
plots of  $(\Delta\Pi^2)_n$ with  $n=10$
for $A_\oplus=0$ and for $A_\oplus=A_\otimes$, respectively.
Table \ref{table1} and Table \ref{table3} list the increase with $n$
of the amplitudes of extrema $\Delta\Pi^2_{max}$ and $\Delta\Pi^2_{min}$
 for $A_\oplus=0$ and  for $A_\oplus=A_\otimes$, respectively.

\begin{table}
 \caption{\label{table1}
 The case $A_\oplus=0$.
 The amplitudes of extrema of $( \Delta\Pi^2)_n$ increase with $n$. }
\begin{center}
 \begin{tabular}{|l|ll|ll|}\hline
 n&$\varpi_{\rm{max}}$&$
    \Delta\Pi^2_{\rm{max}}/A_\otimes$&$\varpi_{\rm{min}}
              $&$\Delta\Pi^2_{\rm{min}}/A_\otimes$ \\
 \hline
1&1.434&1.864&0.743&-0.643\\
\hline
10&1.038&12.027&0.964&-10.761\\
\hline
 100&1.0037&114.456&0.9963&-113.189\\
 \hline
 1000&1.00037&1138.85&0.99963&-1137.58\\
 \hline
 2000&1.00019&2277.07&0.99982&-2275.8\\
 \hline
 10000&1.00004&11382.8&0.999963&-11381.5\\
 \hline
\end{tabular}
\end{center}
\end{table}

\begin{table}
\caption{\label{table3}
The case  $A_\oplus=A_\otimes$.
The amplitudes of extrema of $( \Delta\Pi^2)_n$ increase with $n$.}
\begin{center}
 \begin{tabular}{|l|ll|ll|}\hline
 n&$\varpi_{\rm{max}}
   $&$\Delta\Pi^2_{\rm{max}}/A_\otimes$&$\varpi_{\rm{min}}
                  $&$\Delta\Pi^2_{\rm{min}}/A_\otimes$ \\
 \hline
1&1.546&1.842&0.889&-1.802\\
\hline
10&1.055&11.05&0.982&-20.214\\
\hline
 100&1.0055&103.504&0.9981&-205.148\\
 \hline
 1000&1.00055&1028.1&0.99981&-2054.59\\
 \hline
 2000&1.00027&2055.43&0.999907&-4109.52\\
 \hline
 10000&1.00005&10274.1&0.999981&-20549\\
 \hline
 \end{tabular}
\end{center}
\end{table}

For circularly polarized GWs in the $x^3-$ direction,
the $n-$ cycle result is
 \be
(\Delta\Pi^2)_n=\frac{A\varpi\sin(2\pi n\varpi)}{2(1+\varpi)},
 \ee
which  does not accumulate with $n$,
but vibrates more rapidly than Eq.(\ref{circulargw}).

Secondly, for the lineal polarized incident GWs in the $x^1-$ direction,
the $n-$ cycle result is
 \be
  (\Delta\Pi^2)_{n}=A_{\otimes}\sin^2{(n\pi \varpi)},
 \ee
which has no  accumulating  effect.
For the circularly polarized GWs in the $x^1-$ direction
 \be
 (\Delta\Pi^2)_n=\frac{A\sin(2n\pi\varpi)}{2},
 \ee
having no accumulating effect, either.

So the above detailed analysis on the $n$-cycle accumulating effects
yields the simple conclusion:
Only linearly polarized incident GWs  in the $x^3$-axis
has a linearly accumulating effect of rotation of PV of EWs
in the limit $\varpi \rightarrow 1 $.
Theoretically,
in order to experimentally obtain
a maximum effect of $n$-cycle accumulation,
the circling EWs in the waveguide
should be running so that $n$ is as large as possible.
Of course, due to attenuation of EWs in the actual waveguide,
for a given waveguide made of conducting metal, such as copper,
an input beam of EWs in the waveguide
can run only a finite number of turns around the torus.
The maximum value of $n$ is approximately equal to
the quality factor $Q$,
mainly determined by the conducting metal employed
and the pump resonances.
For instance,
Cruise's group \cite{cruise3} has used copper for the waveguide,
and the measured value of the quality factor $Q\simeq 2000$.
As for the selective response of the detector to
the particular $x^3-$ direction of the incident GWs
under the resonant condition $\omega \simeq \omega_0 $,
it is a problem for
gravitational radiations from certain sources,
since they generally exist for a finite short period of time
(from minutes to hours)
and have some fixed direction of propagation.
But for the RGWs as the detection object,
it is not a problem at all,
as they consist of various modes in all directions and of
all frequencies, moreover,
they are a stochastic background existing
everywhere and all the time.
Therefore, the RGWs serve as a natural object of detection.
What one needs to do is to set up a convenient position
of the torus and
to fix the the cycling angular frequency $\omega_0 = v/R$
of EWs around the waveguide.
There are always modes of the RGWs with the $x^3$ direction and
the angular frequency $\omega \simeq \omega_0 $.

\section { Detecting capability for very high frequency RGW}

Let us examine the capability of the waveguide detector
built up by Cruise's group \cite{cruise3},
particularly in regards  to the RGWs.
Consider the favorable case of GWs travelling along the $x^3-$direction.
Since the rotation
$\Delta\Pi^2$ is small,
it is equal to the angle $\alpha$ rotated,
$\alpha  \simeq \Delta\Pi^2$.
This angle can be measured by the electric probe.
In general, the  detector sensitivity
 will be limited by the thermal noise in the electronic amplifiers.
 It has been found that \cite{cruise3}
the minimum detectable angle of rotation
\be
\alpha_{\rm{min}}=\sqrt{\frac{ab\, kT B}{fPl^2}},
\ee
 where $f$ is an efficiency factor of the probe,
 which  transfers electric signals
 to the following electronic amplifiers,
 $k$ is the Boltzmann's constant,
 $T$ is the amplifier noise temperature,
and $B$ is the detector bandwidth in hertz.
Thus, for a constant amplitude on the time scale $\sim Q/\nu_0$,
during which the EWs travel  $Q$ turns around the loop,
by Eq.(\ref{Deltan2}),
the minimum detectable amplitude $h_{min}$ of the GWs
 is
\be\label{sencitivity}
\label{hmin} h_{min} = \frac{2}{\pi}\frac{\alpha_{min}}{Q}
=\frac{2}{\pi}\sqrt{\frac{abkTB}{fP_{\rm{in}}Q^3 l^2}},
 \ee
where the input power $P_{in}$ is related
to the circulating power $P$ by $ P_{in}=P/Q$.
Here $Q$ is the quality factor of the waveguide.
For a random signals of GWs with amplitude varying
considerably over the time scale $\sim Q/\nu_0$,
the minimum detectable amplitude is
 \be  \label{h2}
h_{\rm{min}}=\frac{2}{\pi}\sqrt{\frac{abkTB}{fP_{\rm{in}}Q^2l^2}},
 \ee
since the angle $\alpha$ of rotation accumulatively
increases  as $\alpha \propto\sqrt{Q}$ as for a random walk.

\begin{figure}
\centerline{\includegraphics[width=8cm]{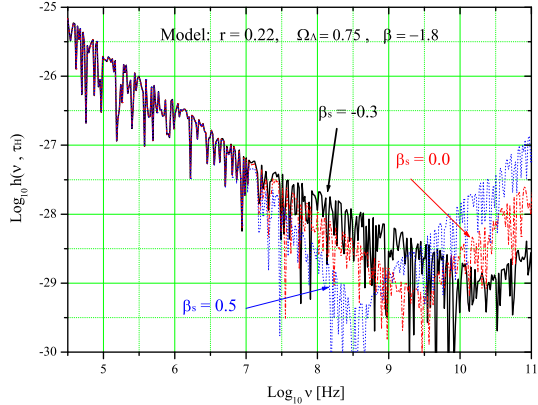}}
\caption{\label{fig10} The spectrum $h(\nu,\eta_{H})$ with
different parameter $\beta_{s}$ for $\beta=-1.8$.}
\end{figure}

The waveguide detector is used to monitor the GWs
of frequency $\sim 10^8$ Hz,
which primarily come from the stochastic background of RGWs
with a very broad frequency range ($10^{-18} \sim 10^{10} $) Hz
\cite{Zhang, Miao, Grishchuk2}.
The RGW spectrum $h(\nu,\eta_{H})$ as given by Eq.(\ref{spectrum})
in a frequency range $>10^7 Hz$
depends sensitively on the reheating parameter $\beta_{s}$.
The spectra for three different values of
$\beta_{s}=0.5,\, 0,\,-0.3$, respectively,
are given in Fig.\ref{fig10}
for a model  with $\beta=-1.8$, $r=0.22$, and $\Omega_{A}=0.75$.
A larger $\beta_{s}$ has
a lower amplitude in the range $(10^7\sim 10^9)$ Hz,
however, around $\nu  \ge  10^9$ Hz
the spectrum begins to increase considerably.
Therefore, if the detector is capable of accurately
detecting the RGWs signals,
it will, in principle,  be able to
constrain the model parameters  $\beta$ and $\beta_{s}$,
and distinguish different models of reheating during the early universe.

Other sources of GWs, such as binary neutron stars or black holes,
merging of neutron stars or black holes can produce GWs,
but the frequency is much lower than $ 10^{8} $ Hz \cite{zhang2}.
Thus they are not to be detected by the waveguide detector
discussed here.
There might be other astrophysical processes,
which can give rise to high frequency GWs \cite{Bisnovatyi}.
Thermal gravitational radiation of stars can generate GWs
at most probable frequencies $\sim 10^{15}$ Hz.
But this frequency is too high for the waveguide detector
\cite{Bisnovatyi} \cite{Weinberg}.
The predicted graser beams in interstellar plasma
can generate GWs with ``optical'' frequencies $> 10^{12}$ Hz,
still too high for the waveguide detector
\cite{Servin}.
The GW radiation from primordial black holes
of small mass $M\le 10^{-5} M_\odot$
can generate GWs with frequencies  $\ge 10^{10}$ Hz,
which is too high for the waveguide detector.
Besides the rate of event is very
low $\sim 5\times 10^{-2}$events/year/galaxy
 \cite{nakamura}.
Therefore, our primary object of detection is RGWs, whose spectrum
in the current accelerating universe is derived in
Refs.\cite{Zhang,Miao}.

What the waveguide detector actually detects is the
root-mean-square (r.m.s.) amplitude  of RGWs  per Hz$ ^{1/2}$
 at a given $\nu$,
which  can be written simply as~\cite{Grishchuk2}
 \be\label{rms}
 \frac{h(\nu)}{\sqrt{\nu}},
\ee
where  $h(\nu)$ denotes the value of
the spectrum $h(\nu,\eta_{H})$ given in Eq.(\ref{spectrum}).
Since the waveguide detector works around the frequency $10^8$ Hz,
so we need to examine the value of $h(\nu,\eta_{H})$
around this frequency predicted by our calculation \cite{Zhang, Miao}.
For a cosmological model with the tensor/scalar ratio $r=0.22$,
the dark energy $\Omega_\Lambda =0.75$,
and the reheating $\beta_s=0.3$,
one directly reads from Fig.\ref{fig1} the values
$h(\nu) \simeq 10^{-28}, 10^{-34}$ for the values
of the inflationary parameter
$\beta=-1.8$, $ -2.02$, respectively.
So the corresponding  r.m.s amplitude per $Hz^{1/2}$ at
$\nu=10^8 Hz$ is then
\be   \label{rms}
 \frac{h(\nu)}{\sqrt{\nu}} \simeq  (10^{-32} ,
  \,\,\,\, 10^{-37}) \,\,  Hz^{-1/2},
\ee
for the two values of $\beta$, respectively.
On the other hand,
the detector sensitivity can be improved by using
the cross correlation of two or multiple detectors.
From a short run of some $4$ seconds of the two detectors,
Ref.\cite{cruise3} gives the cross correlation sensitivity,
 \be   \label{sqrt}
 5\times10^{-15} Hz^{-1/2},
 \ee
 which is within a factor $4$ of the predicted
sensitivity given that parameters $P_{in}=69 mW$, $T=300$K,
$Q=2000$, $(ad)/l^2=0.5$, and $f>0.9$.
By comparing the preliminary experimental
result in Eq.(\ref{sqrt})
with the predicted values in Eq.(\ref{rms}),
it is clear that the predicted value of RGWs
in the model $\beta=-1.8$
is lower than the prototype detector sensitivity by  $17$ orders.
As has been analyzed in Ref.\cite{cruise3},
the detector sensitivity of the current detector
could be improved by a factor of $10^4 \sim  10^5$,
through optimization of the transducers,
use of cryogenic amplifiers and multiple detector correlation.
But even with these possible improvements,
still that will be some  $12$ orders short to able to measure
the predicted amplitude of RGWs in Eq.(\ref{rms}).

An interesting feature of the spectrum of RGWs
is that it has a higher amplitude in lower frequencies.
This may be suggestive for new ways of enhancing
the chance of detections.
As is seen from Eq.(\ref{nu-g}),
if one increases the radius $R$ of the annular waveguide,
say to from $1$ meter to $100$ meters,
the frequency of GWs to be detected will subsequently be reduced
to a low value $\nu_g\simeq 5\times 10^5$ Hz,
at which the spectral amplitude $h(\nu,\eta_H)$
increases by a factor $\sim 10^3$,
as is seen from Fig.\ref{fig1}.
The  r.m.s amplitude per Hz$^{1/2}$
will be $h(\nu)/\sqrt{\nu}\sim  10^{-28}$ Hz$^{-1/2}$.
Now, this is only some $8$ orders lower than
the detector sensitivity of the improved device.
Therefore, according to our calculation of RGWs,
it is unlikely to detect signals of RGWs,
using the annular waveguide detectors as it stands today.
But enlarging the radius $R$ will enhance the
detection probability considerably, of course,
at the price of a larger sum of cost
and a more complex construction.
By the way,
note that LIGO is still
unable to detect the RGWs by 2 orders of magnitude
even it has achieved its design sensitivity \cite{Miao}.
Moreover, theoretically,
there are  possibilities
that the waveguide detector can detect signals from other kinds
of sources  of GWS with a much improved sensitivity.

\section{ Conclusions}

From the calculations of the rotation of PV of EWs,
it is found that the detector only essentially  responds to
the linearly polarized RGWs travelling in the $x^3$-axis
under the resonant condition.
Both the circularly polarized RGWs travelling
along any direction and the linearly polarized RGWs travelling in
the $x^1$-  or  $x^2$-axis have not observable effect.
But these propose no problem for the RGWs as the object of detection.

From our analysis comparing the spectrum of GRWs
with the detector sensitivity,
the RGWs in the accelerating universe have a very low amplitude
and are not possible to detect using this current detector.
The gap between them is some 17 orders of
magnitude under the current experiment conditions.
Even with the improvements on the current detector system
as planned in \cite{cruise3},
there will still be a gap of $12$ orders of magnitude.
Examining the detector itself,
Eq.(\ref{hmin}) and Eq.(\ref{h2}) tell that
the sensitivity of the detector can be directly
improved by several means as follows:
(1) The use  of cryogenic devices at lower temperature
     $T$ of the environment, i.e., to reduce thermal noise
     of the amplifiers;
(2) Increasing the quality factor $Q$ of the waveguide,
    so that the EWs can travel more number of turns around the loop
    path;
(3) Enhancing the input power $P_{in}$ of the EWs into the waveguide;
(4) Using  multiple detectors, whose correlation can
         improve the sensitivity of the detector.
On the other hand,
the shape of the  RGWs spectrum $h(\nu)$ is such that
its amplitude is higher in lower frequencies.
Therefore,  it may be more promising to detect
the  RGWs  in the  relative lower frequency range.
For instance, if the radius of torus is increased to $R=100$ meter,
the detecting frequency $\nu_g \sim 5\times 10^5$ Hz,
and the gap will reduced down to $8$ orders of magnitude.
An overall estimate is that
significant improvements of the current prototype detector are needed
for a possible detection of RGWs by the waveguide.

ACKNOLEDGMENT: We thank  Dr. A.M. Cruise for helpful suggestions.
Y.Zhang's work was supported by the CNSF No.10173008,
 SRFDP, and CAS.

\end{document}